# X-Ray Analysis of an Off-axis Merger Stage Binary Galaxy Cluster: PSZ2 G279.79+39.09

Sibel Döner[1,2], Turgay Caglar[2,3], Krista L. Smith[2], Serap Ak[4], Andrea Botteon[5], M. Kiyami Erdim[6], and John A. ZuHone[7]

[1] Istanbul University, Institute of Graduate Studies in Science, Programme of Astronomy and Space Sciences, 34116 Beyazit, Istanbul, Türkiye; sibeldnr@gmail.com
[2] George P. and Cynthia Woods Mitchell Institute for Fundamental Physics and Astronomy, Texas A&M University, College Station, TX 77843-4242, USA
[3] Leiden Observatory, PO Box 9513, 2300 RA Leiden, The Netherlands
[4] Istanbul University, Faculty of Science, Department of Astronomy and Space Sciences, 34116 Istanbul, Türkiye
[5] INAF—IRA, Via Piero Gobetti 101, 40129 Bologna, Italy
[6] Yildiz Technical University, Faculty of Arts and Science, Department of Physics, Istanbul 34220, Türkiye
[7] Center for Astrophysics |Harvard & Smithsonian, 60 Garden Street, Cambridge, MA 02138, USA

## Abstract

We present an X-ray analysis of the merging galaxy cluster system PSZ2 G279.79+39.09 ($z = 0.29$) using archival XMM-Newton and Chandra observations. The surface brightness image is bimodal, elongated east–west with a projected core separation of ∼1.35 Mpc. We measure gas temperatures of 5.36 keV for the eastern subcluster (PSZ-E) and 5.44 keV for the western component (PSZ-W). Assuming isothermal intracluster gas, the hydrostatic masses are $\log(M_{500}/M_{\odot}) = 14.76$ for PSZ-E and 14.54 for PSZ-W, implying a mass ratio of ∼1:1.7. PSZ-E shows X-ray concentration indices of $c_{40}/c_{400} = 0.124$ and $c_{100}/c_{500} = 0.278$, together with a centroid shift of $w = 0.016$, indicating a disturbed halo that still hosts a compact cool core; PSZ-W is comparably disturbed even in its core. Both subclusters exhibit intracluster medium asymmetries consistent with ram-pressure stripping, and PSZ-W displays an X-ray tail extending nearly to PSZ-E's outskirts. The orientation and length of this tail support an off-axis merger geometry. Thermodynamic maps reveal a hot (∼7.3 keV), high-pressure, high-entropy bridge between the cores. From the Rankine–Hugoniot temperature jump, we infer a Mach number $M = 1.48^{+0.30}_{-0.28}$, consistent with a weak merger shock propagating at $1695^{+316}_{-300}$ [km s$^{-1}$. These results indicate a merger with a nonzero impact parameter, likely observed near core passage ($\lesssim$0.5 Gyr before or after), with the prepericenter scenario slightly preferred based on the projected separation and thermodynamic structure.

## 1. Introduction

Galaxy clusters are the most massive gravitationally bound systems in the Universe, hosting hundreds to thousands of galaxies embedded within deep gravitational potential wells. They assemble hierarchically through successive mergers of smaller substructures, making them exceptional laboratories for investigating some of the most energetic events in the cosmos (releasing up to $\sim 10^{64}$ erg) via X-ray observations of the hot intracluster medium (ICM; C. L. Sarazin 2002). Studying these mergers is essential for understanding the thermodynamics of the ICM and for tracing the formation and evolution of large-scale structure in the Universe.

Previous X-ray studies have examined a wide range of merging galaxy cluster systems, revealing a rich set of dynamical features (see M. Markevitch & A. Vikhlinin 2007; L. Feretti et al. 2012, and references therein). X-ray observations are compelling for studying cluster mergers, as they allow us to trace the behavior of the hot, diffuse ICM, which responds strongly to gravitational interactions (e.g., H. Böhringer & N. Werner 2010; A. Botteon et al. 2019). Variations in temperature, pressure, density, and entropy—key indicators of ICM thermodynamics—can be directly probed through these observations, providing insight into the heating mechanisms at play (I. G. McCarthy et al. 2007; J. A. ZuHone et al. 2010; J. A. ZuHone 2011; H. Bourdin et al. 2013; X. Shi et al. 2020). Moreover, X-ray studies have revealed a range of characteristic signatures associated with mergers, including shock fronts, cold fronts, turbulence, ram-pressure stripping, and adiabatic compression (for a review, see M. Markevitch & A. Vikhlinin 2007). Revealing the composition and thermal properties of the ICM can help us to identify the merging stage of a galaxy cluster and gain deeper insight into how such cosmic structures evolve. Galaxy cluster mergers can be roughly divided into two states, characterized by distinct features: pre- and postcore passage. Before the core passage, distinct X-ray centroids of merging structures (such as X-ray gas clumps) are observed. Additionally, temperature asymmetries, surface brightness (SB) discontinuities, weak X-ray shocks, and diffuse X-ray tails pointing along the collision axis can also be present (T. Caglar & M. Hudaverdi 2017; W. R. Forman 2017; A. Botteon et al. 2018; E. J. Hallman et al. 2018; L. Gu et al. 2019; H. I. Kaya et al. 2019; K. Migkas et al. 2025; A. Stroe et al. 2025). The orientation of merger-driven shocks depends not only on the merger geometry (e.g., head on versus off axis) but also on the merger stage; such shocks can appear between the subclusters (axial) or perpendicular to the merger axis (equatorial), even in head-on mergers after core passage (J.-H. Ha et al. 2018). After the core passage, X-ray analysis of temperature, entropy, and pressure distributions is more critical to diagnose the dynamical state. Sharp discontinuities in SB and temperature can be present, caused by the outward propagation of weak shocks and turbulence in the gas, and all these signatures might be remnants of the previous core collision (M. Markevitch & A. Vikhlinin 2007). Many hydrodynamic simulations of cluster mergers have also shown that turbulence and gas sloshing, triggered by cluster mergers, can persist long after the core passage and significantly influence the thermal and dynamic structure of the ICM (J. A. ZuHone et al. 2010, 2011; E. Roediger et al. 2011; R. E. G. Machado et al. 2022).

There are numerous studies concentrated on later stages of merging galaxy clusters, while X-ray investigations of mergers where the clusters are close to core passage are still limited to a couple of dozen (e.g., T. Caglar 2018, and the references therein). To fully capture the physics of earlier stages of mergers, more detailed X-ray investigations are necessary. Interestingly, most signatures attributed to postpericenter mergers are also present in some prepericenter systems, such as X-ray shocks, ICM asymmetries, ram-pressure stripped gas, and radio halos; however, they are reported to be relatively weaker (H. Ebeling et al. 2017; A. Botteon et al. 2018).

In merging galaxy clusters, low-SB X-ray bridges are often interpreted as signatures of early stage interactions where the ICM becomes compressed between the approaching subclusters (e.g., L. Gu et al. 2019). Such bridges have been observed in systems such as A399-A401 (F. Govoni et al. 2019) and A1758 (A. Botteon et al. 2020). These structures can arise from weak shocks, ram-pressure compression, or adiabatic heating of the gas between the subclusters, and they provide strong diagnostic power for identifying the merger stage. The presence of a diffuse X-ray bridge in PSZ2 G279.79+39.09 coincident with faint radio emission (A. Botteon et al. 2026) makes this system a promising example of a pre- or near-pericenter off-axis merger where bridge formation precedes strong shock heating.

We aim to perform a detailed X-ray investigation of a merging galaxy cluster candidate, PSZ2 G279.79+39.09, recently detected by the Planck Survey (R.A. = 173.°88, decl. = –20.°3403, $z = 0.29$, Galactic longitude = 279.°794, Galactic latitude = 39.°0963) for the first time in the literature (Planck Collaboration et al. 2014, 2016). The dynamic processes of PSZ2 G279.79+39.09 triggered by mergers are discussed in this work, while in another article we focus on the radio analysis of the system (see A. Botteon et al. 2026). The paper is organized as follows: Section 2 presents the observation logs and data processing; Section 3 describes the X-ray data analysis methods; Section 4 covers the discussion of our results; and finally, we summarize our findings in Section 5. In this paper, we adopt standard ΛCDM cosmology parameters: $H_0 = 70\,\mathrm{km\,s^{-1}\,Mpc^{-1}}$, $\Omega_M = 0.3$, and $\Omega_\Lambda = 0.7$ for a flat Universe. In this cosmology, $1''$ corresponds to 4.350 kpc for $z = 0.29$. Unless stated otherwise, the error values are quoted in the 90% confidence interval in our analysis.

## 2. Observation and Data Reduction

The XMM-Newton data were acquired on 2014 June 13 for a total exposure time of 43.4 ks with the observation ID 0744390101. Thin filters were used for the EPIC-MOS1, EPIC-MOS2, and EPIC-PN detectors. X-ray observations were taken in full frame for MOS1 and MOS2 and extended full frame for PN. The Chandra observation was performed on 2015 April 25 with an effective exposure time of 24.73 ks in VFAINT mode with the observation ID 16128. The X-ray observational data were taken from the XMM-Newton Science Archive and the Chandra Data Archive (CDA). We used XMM-Newton Science Analysis Software (XMM-SAS v.22.1.0) and XMM-Newton Extended Source Analysis Software (XMM-ESAS), and Chandra Interactive Analysis of Observations (CIAO v.4.17) to reduce the XMM-Newton and Chandra X-ray data, respectively.

The data files from XMM-Newton were processed as follows. The calibration files were created using the `cifbuild` task, and the observation summary file was generated through the `odfingest` task. The calibrated event lists for the observation were created using the `emchain` (EPIC-MOSs) and `epchain` (EPIC-PN) processes. The X-ray light curve was generated separately for each of the MOS1, MOS2, and PN cameras with the `evselect` command and later used for data cleaning with the data-clipping method. After filtering for high-background intervals, the net good-time exposures were 27.4 ks for each MOS camera and 10 ks for the PN camera. The point sources in the field of view are detected using the `edetect_chain` task in five different bands within the 0.2–12 keV energy range and they are subtracted from the data. Finally, the vignetting-corrected background-subtracted XMM-Newton image was generated using the `eimageget` and `eimagecombine` tasks for only the MOS data, since the PN camera has many CCD gaps affecting the SB profile (see the left panel of Figure 1).

For processing Chandra data files, the Chandra observation number 16128 was downloaded from the CDA and analyzed using CIAO. The data were reprocessed using the `chandra_repro` task to generate new level 2 data files. The X-ray light curve was generated using `dmextract`, and the bad data outside 3$\sigma$ of the mean were removed using the `lc_clean` task, resulting in a net good-time exposure of 23.0 ks. Point sources are detected using `wavdetect`. A background event file is created using the `blanksky` task. Raw and background-subtracted X-ray images were generated using the `fluximage` and `blanksky_image` tasks, respectively (see the right panel of Figure 1). Finally, we used the `specextract` task to generate the spectral files. However, after careful consideration, we decided to use the Chandra data only for spatial analysis, as the available counts are insufficient for reliable spectral fitting, with ∼2000 total counts in the western subcluster (hereafter, PSZ-W) and ∼2700 total counts in the eastern subcluster (hereafter, PSZ-E).

For the XMM-Newton data, we perform the spectral data processing and background modeling with XMM-ESAS. After preparing the data with the `emanom`, `espfilt`, and `cheese` tasks, we used the `mos-spectra` and `pn-spectra` tasks to generate spectra, redistribution matrix files, and ancillary response files. For instrumental background generation, we used the `mos-back` and `pn-back` tasks. The X-ray background was modeled via fitting solar wind charge exchange, soft proton, and cosmic background components. Further details about X-ray background modeling can be found in J. Nevalainen et al. (2005) and S. L. Snowden et al. (2008). Since background model components are complicated and background modeling is time-consuming, we also extract spectral files using XMM-SAS. Therefore, the `evselect` task is used to generate both source and background files,[8] the `rmfgen` task is used for redistribution matrix files, and the `arfgen` task is used to generate ancillary response files. For both XMM-ESAS and XMM-SAS, we extracted spectral files from a circular region with a radius of 2$'$.5 for PSZ-E and 2$'$.25 for PSZ-W, with radii selected to approximately match the extent where the X-ray SB drops to the background level (see Figure 1). We extract local background data from the dashed annulus around the cluster pair within an inner radius of 10$'$.5 and an outer radius of 11$'$.

[8] Unless otherwise stated, `FLAG` == 0 was applied for both MOS and PN data, with `PATTERN` ⩽ 12 for MOS and `PATTERN` ⩽ 4 for PN.

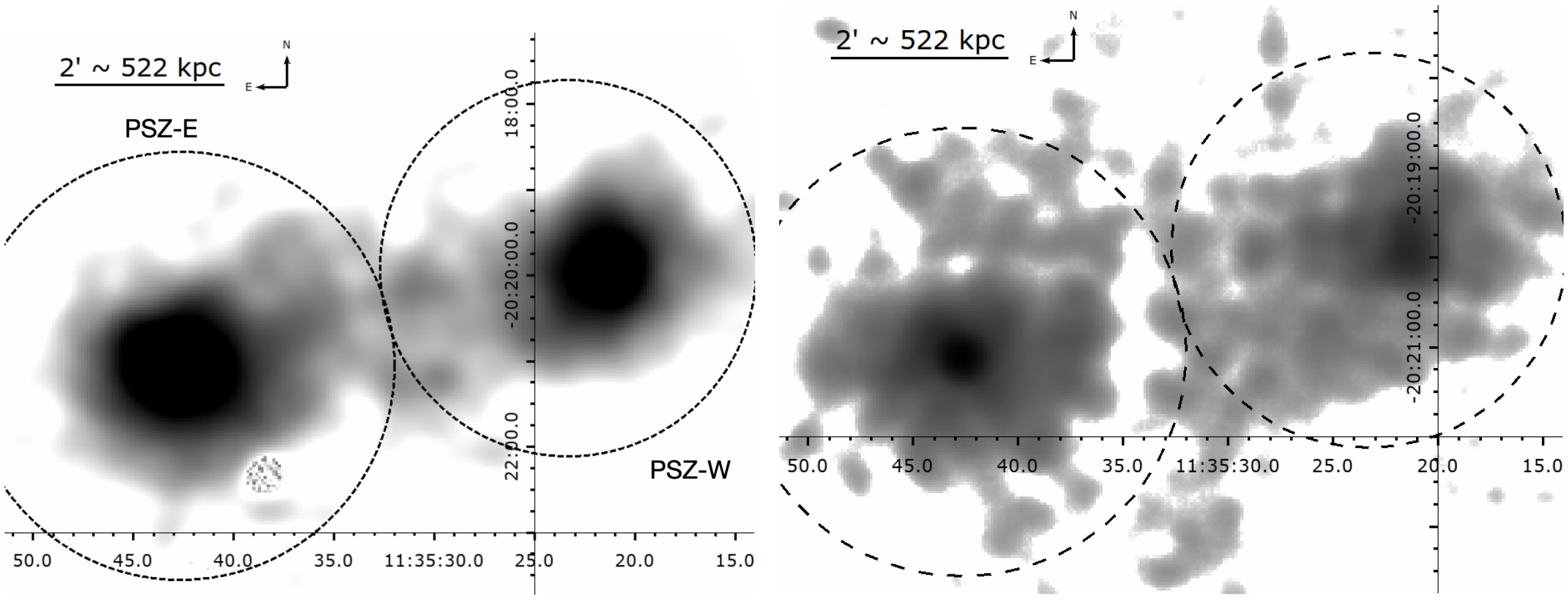

**Figure 1.** Left: the adaptively smoothed vignetting-corrected background-subtracted XMM-Newton image in the 0.5–2.0 keV band. Right: the 4.5$\sigma$ smoothed background-subtracted Chandra image in the 0.5–2.0 keV band. The dashed circles represent each subcluster's ICM, which will later be used to obtain the mean cluster spectral properties. The apparent differences in morphology between the XMM-Newton and Chandra images are due to instrumental effects. The Chandra image (ACIS-I) has finer angular resolution and smaller smoothing, revealing compact features and apparent discontinuities where CCD gaps intersect diffuse emission, while the broader point-spread function (PSF) and adaptive smoothing in XMM-Newton produce a smoother, more symmetric appearance.

## 3. Analysis

### 3.1. Subclusters' Mean X-Ray Outputs

The XMM-Newton spectra of our subclusters were fitted using an emission spectrum from collisionally ionized diffuse gas (APEC; R. K. Smith et al. 2001) together with a photoelectric absorption model using cross sections (PHABS) from the XSPEC software package, v.12.15.0 (K. A. Arnaud 1996). The temperature, abundance, and normalization were allowed to be free parameters. We adopt the solar abundance table from E. Anders & N. Grevesse (1989). The Galactic absorption column density is fixed to $4.30 \times 10^{20}\ \mathrm{cm}^{-2}$ obtained from the Leiden/Argentine/Bonn Survey (P. M. W. Kalberla et al. 2005). We restricted the fitted energy band to 0.3–10 keV, simultaneously modeling the EPIC-MOS and PN spectra once their consistency had been verified. We excluded prominent high-energy lines from the quiescent particle background, namely Cu K$\alpha$ at 8.04 keV for PN, from the spectral fits, by omitting the 7.9–8.1 keV energy range. Throughout this paper, we use the C-statistic to evaluate the goodness of fit of our models; it yields more reliable results for low-count data without introducing bias in the derived parameters (W. Cash 1979).

For XMM-ESAS, we found an X-ray temperature ($kT$) of $5.29 \pm 0.17$ keV and an abundance ($Z$) of $0.32 \pm 0.05\ Z_\odot$ for PSZ-E and $kT = 5.79 \pm 0.77$ keV and $Z = 0.25 \pm 0.09\ Z_\odot$ for PSZ-W. For XMM-SAS, we found $kT = 5.36 \pm 0.15$ keV and $Z = 0.34 \pm 0.04\ Z_\odot$ for PSZ-E and $kT = 5.44 \pm 0.65$ keV and $Z = 0.29 \pm 0.07\ Z_\odot$ for PSZ-W. Since both results from XMM-ESAS and XMM-SAS are consistent with each other, we decided to proceed with XMM-SAS for the rest of our analysis.

### 3.2. The X-Ray and Optical Centroid Separation

We computed the projected offset between the X-ray and optical centroids to identify possible merger-induced disturbances in the cluster core. The separation is given by

$$\Delta_\mathrm{r} = \sqrt{[(\alpha_\mathrm{opt} - \alpha_\mathrm{X})\cos\bar{\delta}]^2 + (\delta_\mathrm{opt} - \delta_\mathrm{X})^2} \times c, \qquad (1)$$

where $\bar{\delta} = (\delta_\mathrm{opt} + \delta_\mathrm{X})/2$ and $c$ is the angular scale in kiloparsecs per arcsecond. The X-ray centroid ($\alpha_\mathrm{X}$, $\delta_\mathrm{X}$) was derived from the photon-weighted center within the soft-band (0.5–2 keV) X-ray image, while the optical centroid ($\alpha_\mathrm{opt}$, $\delta_\mathrm{opt}$) corresponds to the location of the brightest cluster galaxy (BCG), identified using deep $r$-band imaging. The photon-weighted center was computed using an aperture corresponding to a single ACIS-I pixel size ($\sim$0$''$.5). The uncertainties on the X-ray–optical centroid offset were estimated through a bootstrap Monte Carlo approach using 1,000,000 realizations. In each iteration, the R.A. and decl. of both centroids were perturbed by random Gaussian deviations consistent with their positional uncertainties (0$''$.5 for the X-ray centroid and 0$''$.2 for the optical centroid), and the separation was recalculated. The 1$\sigma$ uncertainties were then derived from the 16th and 84th percentiles of the resulting offset distribution. All computed offsets are reported in Table 1, along with the X-ray to optical association probability $P_{\mathrm{X/O}}$, which was estimated using Equation (2) of F. X. Pineau et al. (2017)

### 3.3. The X-Ray Surface Brightness Profile

The X-ray SB, defined as the line-of-sight projection of the plasma emissivity per unit solid angle, was modeled for each substructure using the isothermal $\beta$-profile (A. Cavaliere & R. Fusco-Femiano 1976). Specifically, we fit

$$S(r) = S_0\,[1 + (r/r_\mathrm{c})^2]^{-3\beta+\frac{1}{2}} + C, \qquad (2)$$

**Table 1**
Derived X-Ray and Mass-related Parameters for the Eastern and Western Components of PSZ2 G279.79+39.09

| Parameter | PSZ-E | PSZ-W |
|---|---|---|
| $kT$ (keV) | $5.36 \pm 0.14$ | $5.44 \pm 0.65$ |
| $Z$ ($Z_\odot$) | $0.34 \pm 0.04$ | $0.29 \pm\ .07$ |
| $L_X$ (erg s$^{-1}$) | $(2.97 \pm 0.04) \times 10^{44}$ | $(2.05 \pm 0.05) \times 10^{44}$ |
| $\Delta r$ (kpc) | $6.67 \pm 2.20$ | $5.67 \pm 2.05$ |
| $P_{X/O}$ | 92% | 91% |
| $r_c$[a] (kpc) | $182 \pm 24$ | $75 \pm 15$ |
| $\beta$[a] | $0.68 \pm 0.08$ | $0.47 \pm 0.04$ |
| $r_{500}$ (kpc) | $1029.9^{+69.2}_{-70.5}$ | $862.6^{+62.6}_{-63.7}$ |
| $M_{500}$ ($M_\odot$) | $(4.05^{+0.89}_{-0.79}) \times 10^{14}$ | $(2.44^{+0.57}_{-0.50}) \times 10^{14}$ |
| $M_{500}^{(M-T)}$ ($M_\odot$) | $(3.83 \pm 0.85) \times 10^{14}$ | $(3.92 \pm 1.15) \times 10^{14}$ |
| $n_{e,0}$ (cm$^{-3}$) | $(5.92^{+1.80}_{-1.28}) \times 10^{-3}$ | $(1.06^{+0.40}_{-0.25}) \times 10^{-2}$ |
| $M_{\rm gas}(500)$ ($M_\odot$) | $(5.27^{+0.73}_{-0.74}) \times 10^{13}$ | $(4.76^{+0.72}_{-0.69}) \times 10^{13}$ |
| $f = M_{\rm gas}/M_{500}$ | $0.130^{+0.043}_{-0.035}$ | $0.195^{+0.042}_{-0.037}$ |
| $w$ | $0.016 \pm 0.003$ | $0.039 \pm 0.005$ |
| $P_2/P_0$ | $(7.062 \pm 0.908) \times 10^{-6}$ | $(2.330 \pm 0.1988) \times 10^{-5}$ |
| $P_3/P_0$ | $(2.851 \pm 0.810) \times 10^{-7}$ | $(1.204 \pm 0.576) \times 10^{-7}$ |
| $P_4/P_0$ | $(1.228 \pm 0.414) \times 10^{-7}$ | $(6.231 \pm 8.333) \times 10^{-9}$ |
| $c_{40}/c_{400}$ | $0.124 \pm 0.001$ | $0.062 \pm 0.001$ |
| $c_{100}/c_{500}$ | $0.274 \pm 0.001$ | $0.213 \pm 0.002$ |

**Note.**
[a] The PSZ-E subcluster was fitted with double $\beta$ profiles, where the resulting compact core $\beta$ parameters are $r_{c,\rm core} = 63.5 \pm 9$ kpc and $\beta_{\rm core} = 2.0 \pm 0.92$. Apertures used to derive the listed quantities are described in the text.

where $S_0$ is the central SB, $r_c$ is the core radius, $C$ is a constant parameter for the background level, and $\beta$ is the outer slope parameter. The constant $c$ represents the uniform background level and is included to ensure unbiased determinations of $S_0$, $r_c$, and $\beta$.

The SB profiles were extracted within a circular region of $4'$ radius ($\sim$1 Mpc) centered on each subcluster. To avoid contamination, we masked both the X-ray bridge and the emission from the opposing subcluster in each case. This ensured that the fitted profiles reflect the intrinsic structure of each subcluster independently.

### 3.4. Mass Calculation

Galaxy clusters are large systems dominated by dark matter in their total mass, followed by hot cluster gas and galaxies. X-ray observations are a key method for estimating the gas mass and total mass of these clusters, particularly by analyzing the hot ICM, which emits X-rays due to thermal bremsstrahlung and line emission. First, we start by calculating the dimensionless Hubble parameter to account for the redshift dependence of the cosmic expansion rate, which properly scales both the critical density and the angular diameter distance with redshift $z$:

$$E(z) = \sqrt{\Omega_m(1+z)^3 + \Omega_k(1+z)^2 + \Omega_\Lambda}\,, \qquad (3)$$

where $\Omega_m$ is the matter density parameter, $\Omega_\Lambda$ is the dark-energy density parameter, $\Omega_k = 1 - \Omega_m - \Omega_\Lambda$ is the curvature density parameter, and $z$ is the redshift. Under the assumption that gas in galaxy clusters is considered to be roughly isothermal and follows the $\beta$-model, the characteristic radius $r_\Delta$ can be estimated using the following calculation (G. B. Lima Neto et al. 2003):

$$r_\Delta = r_c \sqrt{\frac{2.3 \times 10^8 \beta kT}{\Delta h_{70}^2 E(z)^2 \mu r_c^2}}\,, \qquad (4)$$

where $\beta$ is the slope parameter of the $\beta$-model obtained by fitting the SB profile, $r_c$ is the characteristic radius in kpc units, and $kT$ is the X-ray temperature in keV units. Under the assumptions of isothermal gas and hydrostatic equilibrium, the total mass of a galaxy cluster can be calculated from the observable X-ray properties:

$$M(r) = \frac{3kT_0\beta\gamma_p r_c}{G\mu m_p}\left(\frac{r}{r_c}\right)^3\left[1 + \left(\frac{r}{r_c}\right)^2\right]^{-1-\frac{3}{2}(\gamma_p-1)\beta}. \qquad (5)$$

Here, $k$ is the Boltzmann constant, $G$ is the gravitational constant, $\mu$ is the mean molecular weight of the ICM (typically $\approx$0.6), $\gamma_p$ is the gas polytropic index (which is reported to be one for isothermal gas and $\sim$1.15 for polytropic gas) and $m_p$ is the proton mass.

We estimate the gas mass and central electron density of the ICM under the assumption of an isothermal $\beta$-model (A. Cavaliere & R. Fusco-Femiano 1976), following the formalism of G. B. Lima Neto et al. (2003). Therefore, $n_e$ can be calculated using the normalization from the APEC model, which is based on the emission measure derived from the X-ray spectra. The $n_e$ is defined as:

$$N = \frac{10^{-14}}{4\pi\, D_A^2\,(1+z)^2}\int n_e\, n_{\rm H}\, dV \ {\rm cm}^{-5}, \qquad (6)$$

where $N$ is the normalization parameter of the APEC model, $D_A$ is the angular distance in units of cm, $n_e$ is the electron density in units of cm$^{-3}$, and $n_{\rm H}$ is the hydrogen number density in units of cm$^{-3}$. For our electron density calculations, we assumed an $n_e/n_{\rm H}$ ratio of 1.2. Here, we note that, due to limitations in photon statistics, we were unable to extract reliable spectra in multiple radial bins. Ideally, one would derive a radial profile of the APEC normalization and perform a deprojected density fit, but our analysis instead uses a single spectrum extracted from a circular region encompassing the intracluster gas of each subcluster. The resulting APEC normalization was analytically inverted under the $\beta$-model assumption to estimate the central electron density $n_{e,0}$. This computation allows us to measure the gas mass ($M_{\rm gas}$) as follows:

$$M_{\rm gas}(r) = 4\pi\, \mu_e\, m_p \int_0^r n_e(r)\, r^2\, dr, \qquad (7)$$

where $\mu_e$ is the mean molecular weight per electron. Finally, the gas-mass fraction can be found using the following equation:

$$f = \frac{M_{\rm gas}(r_{500})}{M_{\rm tot}(r_{500})}. \qquad (8)$$

To estimate the $1\sigma$ uncertainties on $r_{500}$, $M_{500}$, $n_e$, $M_{\rm gas}(r_{500})$, and $f$, we performed a Monte Carlo propagation of the input parameter errors. We generated $N = 10{,}000$ realizations of the parameter sets $\{kT, \beta, r_c, N, r\}$ by drawing each from a Gaussian distribution centered on its best-fit value with standard deviation equal to its measurement uncertainty.

For each realization, we recalculated all derived quantities using the unchanged computational routines and then took the standard deviation of each resulting distribution as the $1\sigma$ uncertainty on the corresponding output parameter. Here we note that we used the same Markov Chain Monte Carlo chains for all parameters to ensure that each realization of $M_{\rm gas}$ and $M_{500}$ is drawn coherently, thereby preserving their covariance in the estimation of $f_{\rm gas}$. This refinement slightly reduced the uncertainties of $f_{\rm gas}$ ($\sim$20%) compared to those obtained without accounting for the covariance between $M_{\rm gas}$ and $M_{500}$.

To cross check our $M_{500}$ calculations, we also computed them using the mass–temperature relation obtained from galaxy clusters reported by V. Toptun et al. (2025) as follows:

$$\log_{10}\left(\frac{M_{500}}{M_\odot}\right) = (1.65 \pm 0.11)\log_{10}\left(\frac{T_{\rm X}}{1\ {\rm keV}}\right) + (13.38 \pm 0.05). \quad (9)$$

Here, we emphasize that the $M$–$T$ relation provides a straightforward, observationally calibrated way to infer cluster masses from X-ray temperatures, making mass estimates accessible even for modest data. Its power-law form reflects self-similar gravitational collapse and works well on average, but it carries $\sim$15% intrinsic scatter and relies on assumptions of hydrostatic equilibrium, sphericity, and instrument calibration. Consequently, it can be biased for disturbed or merging systems and should ideally be cross checked with morphological diagnostics or independent mass probes.

### 3.5. X-Ray Morphological Parameters

X-ray morphological parameters—such as centroid shift, power ratios, and concentration indices—are fundamental tools for probing the dynamical state of galaxy clusters, as they quantify asymmetries, substructures, and core properties. The centroid-shift parameter traces the variation of the X-ray emission center across apertures and reflects dynamic distortion, expressed by the following equation (J. J. Mohr et al. 1995):

$$w = \left[\frac{1}{N-1}\sum_{i=1}^{N}(\Delta_i - \langle\Delta\rangle)^2\right]^{1/2} \times \frac{1}{R_{\rm ap}}, \quad (10)$$

where $N$ is the number of circular apertures used to compute the centroids, $R_{\rm ap}$ is the aperture radius at 500 kpc, and $\Delta$ is the distance.

For the centroid shift, we computed the standard error of the mean, resulting in a $1\sigma$ uncertainty of $w$. The power-ratio method is an analysis based on multipole moments derived from X-ray SB tracing ellipticities, morphological asymmetries, and dynamical states of galaxy clusters (D. A. Buote & J. C. Tsai 1995). The general equation is as follows:

$$P_m = \frac{1}{2m^2 R_{\rm ap}^{2m}}(a_m^2 + b_m^2), \quad (11)$$

where $m$ is the multipole order and $a_m$ and $b_m$ are the multipole coefficients. In this work we compute the power ratios $P_2/P_0$, $P_3/P_0$, and $P_4/P_0$, which respectively trace the cluster ellipticity, asymmetry, and small-scale substructure. The aperture radius $R_{\rm ap}$ is set to 500 kpc. Further details on the implementation and interpretation of these parameters are provided in Section 4.2. We estimated the $1\sigma$ uncertainties on each power ratio by generating 10,000 bootstrap realizations of the $R_{\rm ap}$ SB map, perturbing each pixel by up to $\pm$10%, and taking the standard deviation of the resulting $P_m/P_0$ distributions.

The X-ray concentration index is a measure of how much X-ray luminosity (or SB) is concentrated in the center of a galaxy cluster compared to its outskirts. High X-ray concentration values typically indicate that clusters may host relaxed and cool cores (J. S. Santos et al. 2008). It can be calculated using the following equation:

$$c_{\rm X} = \frac{S(r < r_{\rm in})}{S(r < r_{\rm out})}, \quad (12)$$

where $S$ is the SB and $r$ is the radius. We employ two aperture combinations, $S_{40}/S_{400}$ (J. S. Santos et al. 2008) and $S_{100}/S_{500}$ (R. Cassano et al. 2010), with radii expressed in kiloparsecs. A full description and justification of these definitions is given in Section 4.2. Finally, we estimated the uncertainty in the X-ray concentration index with the same bootstrapping method applied to power ratios. Together with all these parameters, a robust framework distinguishes between relaxed and disturbed cluster morphologies. We note that all X-ray morphology metrics presented in this work were computed separately for each subcluster, using masks that exclude the bridge region and any emission from the companion halo. Thus, the concentration, centroid-shift, and power-ratio measurements reflect only the intrinsic morphology of each subcluster and are not contaminated by flux from the opposite component.

### 3.6. Temperature, Pressure, and Entropy Maps

X-ray temperature maps are crucial tools for illustrating the temperature distribution of the ICM, providing insights into phenomena such as mergers, shock waves, and turbulence. First, we used a set of Python scripts introduced in A. Niemiec et al. (2023), applying the adaptive circular binning (ACB) algorithm to gather a target number of photons around each pixel. It then determines the pseudospectrum of the source in that region and fits it using a spectral template generated with the optically thin thermal plasma model (APEC) in XSPEC. For our data, we utilized the X-ray images in 18 different energy bands within 0.4–10 keV (0.4, 0.5, 0.6, 0.7, 0.8, 0.9, 1.0, 1.1, 1.2, 1.3, 1.5, 1.8, 2.2, 2.8, 3.5, 4.5, 6.2, 7.0, and 10.0 keV), along with their corresponding exposure and background maps. To achieve this, we employed the APEC model together with a fixed Galactic column density of hydrogen ($4.30 \times 10^{20}$ cm$^{-2}$) and metallicity of 0.3 $Z_\odot$, convolved with the instrumental responses of XMM-Newton, to generate spectral templates integrated over the five selected energy bands as a function of plasma temperature. The further details about this method can be found in A. Niemiec et al. (2023) and A. Botteon et al. (2024).

The algorithm also produce APEC normalization, pressure ($P$), and entropy ($K$) maps. Such maps allow us to trace the thermodynamic variations in the ICM and help us to understand the direct physical mechanism responsible for them.

## 4. Results and Discussion

### 4.1. Subclusters

The X-ray SB map of PSZ2 G279.79+39.09 displays a bimodal morphology, with the emission elongated along the

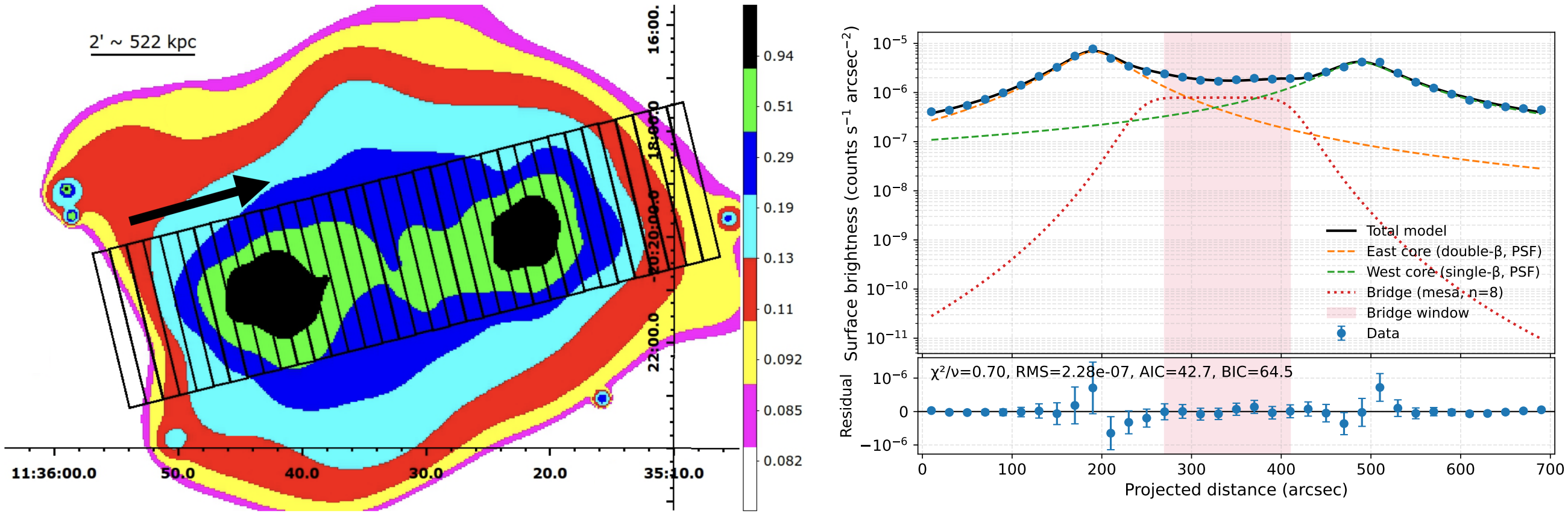

**Figure 2.** Left: the adaptively smoothed vignetting-corrected background-subtracted XMM-Newton image in the 0.5–2.0 keV band demonstrating the selected region to generate the SB profile. Right: PSF-convolved X-ray SB profile along the merger axis. The model (black) combines a double-$\beta$ eastern core, single-$\beta$ western core, and flat-topped `mesa` bridge ($n = 8$).

east–west axis. The eastern subcluster (PSZ-E) extends to a projected radius of $\simeq$1.3 Mpc, whereas the western component (PSZ-W) reaches $\simeq$1.15 Mpc. PSZ-E is brighter in X-rays than PSZ-W, with luminosities $L_X = (2.97 \pm 0.04) \times 10^{44}$ erg s$^{-1}$ and $L_X = (2.05 \pm 0.05) \times 10^{44}$ erg s$^{-1}$, respectively, marking it as the primary cluster (Figure 1). The mean ICM temperatures are statistically consistent, $kT = 5.36 \pm 0.14$ keV for PSZ-E and $kT = 5.44 \pm 0.65$ keV for PSZ-W, and so are the metallicities, $Z = 0.34 \pm 0.04 Z_\odot$ and $0.29 \pm 0.07\, Z_\odot$, respectively. Because both the temperatures and abundances agree within their uncertainties, we cannot rule out the possibility that the two subclusters have experienced similar enrichment and thermodynamic histories before encountering each other.

The X-ray centroids have a projected separation of 1.35 Mpc. Based on a crossmatch with optical DSS images, PSZ-E might be clustered around a possible BCG, WISEA J113542.70-202105.8, and PSZ-W might be clustered around the possible BCG WISEA J113521.35-201956.3. The projected offsets between the X-ray and optical centroids are small for both PSZ-E ($\Delta r = 6.67 \pm 2.20$ kpc, $P_{X/O} = 92\%$) and PSZ-W ($\Delta r = 5.67 \pm 2.05$ kpc, $P_{X/O} = 91\%$), supporting a precore passage phase scenario, as large offsets are typical in strongly disturbed systems, which indicate that the clusters have already crossed pericenter (A. W. Mann & H. Ebeling 2012; R. Seppi et al. 2023).

Both ICMs show asymmetries indicative of ram-pressure stripping; in particular, PSZ-W displays an X-ray tail that extends nearly to the outer edge of PSZ-E. The orientation and length of this tail suggest that the two subclusters are undergoing an off-axis merger rather than a strictly head-on collision (see Figure 1). Nonetheless, the presence of the intercluster bridge, clearly detected in X-rays (and also in radio; see A. Botteon et al. 2026), suggests that the impact parameter of the collision is not very large. As shown in Figure 2, while no sharp discontinuity or SB jump is detected along the merger axis, a broad plateau spanning 270″–410″ is clearly visible between the two subclusters. The mean SB[9] within this region, $S_{\rm bridge} = (1.90 \pm 0.07) \times 10^{-6}$ counts s$^{-1}$ arcsec$^{-2}$, exceeds the outer background level, $S_{\rm bkg} = (4.77 \pm 0.18) \times 10^{-7}$ counts s$^{-1}$ arcsec$^{-2}$, by the excess $\Delta S = (1.42 \pm 0.07) \times 10^{-6}$ counts s$^{-1}$ arcsec$^{-2}$. This corresponds to a 297.8% enhancement ($3.98 \times$ the background) detected at $20.3\sigma$ significance, confirming the presence of excess X-ray emission between the two systems. However, a substantial fraction of this excess arises from the projected outskirts of the two subclusters; therefore, to isolate the intrinsic bridge component from projection effects, we performed a detailed parametric modeling of the SB profile. We modeled the one-dimensional X-ray SB profile extracted along the merger axis using a multicomponent, PSF-convolved fitting procedure (see the right panel of Figure 2). The total model comprises three physically motivated components: a double-$\beta$ profile for the eastern subcluster, a single-$\beta$ profile for the western subcluster, and a flat-topped `mesa` function (with exponent $n = 8$) to describe the diffuse bridge emission. The `mesa` profile provides a flexible representation of broad, low-contrast X-ray emission between merging subclusters and has been successfully applied in similar analyses of intercluster bridges (e.g., M. S. Mirakhor et al. 2022). All components were convolved with the instrumental PSF before fitting. The best-fit model reproduces the data well, with a reduced chi-square of $\chi^2/\nu = 0.70$ and an rms residual of $2.3 \times 10^{-7}$. Model-selection criteria (Akaike information criterion, AIC = 42.7 and Bayesian information criterion, BIC = 64.5) indicate that this three-component configuration provides an optimal balance between fit quality and model complexity. Within the bridge region, the combined emission from the projected outskirts of the two subclusters alone accounts for $\langle S_{\rm core} \rangle = 1.15 \times 10^{-6}$ counts s$^{-1}$ arcsec$^{-2}$, implying that $\sim$61% of the observed plateau originates from projection of the two systems. The modeled bridge component contributes a mean SB of $\langle S_X \rangle = 7.45 \times 10^{-7}$ counts s$^{-1}$ arcsec$^{-2}$ (i.e., $\sim$39% of the total plateau) and a line-integrated brightness of $1.07 \times 10^{-4}$ counts s$^{-1}$ arcsec$^{-1}$.

The resulting $\beta$-model parameters are $r_{\rm c,core} = 63.5 \pm 9$ kpc, $\beta_{\rm core} = 2.0 \pm 0.92$, $r_c = 182 \pm 24$ kpc, and $\beta = 0.68 \pm 0.08$. As a cross check, we refit the SB profile of PSZ-E with a single $\beta$-model, excluding the inner radius of PSZ-E ($r < 75$ kpc), resulting in $r_c = 156 \pm 28$ kpc and $\beta = 0.63 \pm 0.07$. Fitting the PSZ-W SB profile, we also report $r_c = 75 \pm 15$ kpc and $\beta = 0.47 \pm 0.04$. Together, the larger core radius and steeper $\beta$ value for PSZ-E indicate a more extended and diffuse gas distribution compared to PSZ-W, rather than a direct difference in total mass, since both subclusters have comparable temperatures. As demonstrated by earlier studies (e.g., H. Xu et al. 1998; J. S. Santos et al. 2008; N. Ota et al. 2013), cool-core galaxy clusters are best characterized by a double $\beta$-model for their SB profiles. This provides the first indication of the cool-core nature of the PSZ-E subcluster.

[9] The X-ray SB measurements were extracted from background-subtracted, exposure- and vignetting-corrected XMM-Newton MOS1 + MOS2 images in the 0.5–2.0 keV energy band and are expressed in units of counts s$^{-1}$ arcsec$^{-2}$.

Combining the best-fitting $\beta$-model parameters with the spectroscopic APEC temperatures and normalization, we derived the characteristic overdensity radius $r_{500}$, the total mass $M_{500}$, the gas mass within that radius, and the gas-mass fraction $f_{\rm gas} = M_{\rm gas}/M_{\rm tot}$ for each subcluster (Table 1). These masses were calculated under the assumptions of hydrostatic equilibrium and isothermal gas, following the standard $\beta$-model formalism. For PSZ-E, we obtain $r_{500} = 1029.9^{+69.2}_{-70.5}$ Mpc and $M_{500} = (4.05^{+0.89}_{-0.79}) \times 10^{14}\,M_\odot$, confirming that it is the more massive component of the pair. We measure a central electron density of $n_{e,0} \simeq 5.92 \times 10^{-3}\,{\rm cm}^{-3}$ and a gas-mass fraction of $f_{\rm gas} = 0.130^{+0.043}_{-0.036}$, both typical of massive, moderately relaxed clusters at this redshift (S. Ettori et al. 2009; T. F. Laganá et al. 2013). PSZ-W is smaller and less massive ($r_{500} = 862.6^{+62.3}_{-63.7}$ Mpc and $M_{500} = (2.44^{+0.57}_{-0.50}) \times 10^{14}\,M_\odot$), yet it exhibits a higher central density ($n_{e,0} \simeq 1.06 \times 10^{-2}\,{\rm cm}^{-3}$) and a gas-mass fraction of $0.195^{+0.042}_{-0.037}$. As a cross check, we also used the empirical $M$–$T$ relation to estimate $r_{500}$ and $M_{500}$, which are consistent with the values derived assuming isothermal hydrostatic equilibrium (see Table 1). Because our data are not sufficiently deep to perform a reliable spectral deprojection, the central electron densities were instead estimated from the SB $\beta$-model parameters, which may introduce modest biases due to projection effects and unresolved substructures (e.g., F. Vazza et al. 2013; D. Eckert et al. 2015). Consequently, the derived $M_{\rm gas}$ values may be slightly overestimated. The elevated $f_{\rm gas}$ observed in PSZ-W is therefore most likely driven by hydrostatic mass bias caused by nonthermal pressure support, with a possible minor contribution from density normalization uncertainties. Hence, the reported gas-mass fractions should be regarded as upper limits.

We additionally compare our individual subcluster mass estimates with values reported in the literature (P. Tarrío et al. 2019; M. Klein et al. 2023; A. Liu et al. 2024). P. Tarrío et al. (2019) reported a total system mass of $M_{500} = (5.41^{+0.91}_{-1.01}) \times 10^{14}\,M_\odot$ based on the $M_{500}$–$D_A^2 Y_{500}$ scaling relation derived from the Sunyaev–Zel'dovich (SZ) effect. An independent estimate from M. Klein et al. (2023), using the X-ray luminosity–mass ($L_{500}$–$M_{500}$) relation calibrated from the ROSAT all-sky survey, yielded a similar total mass of $M_{500} = 5.01 \times 10^{14} M_\odot$. More recently, A. Liu et al. (2024) reported individual masses of $M_{500} = (6.61 \pm 0.84) \times 10^{14} M_\odot$ for PSZ-E and $M_{500} = (5.64 \pm 0.65) \times 10^{14}\,M_\odot$ for PSZ-W, obtained from the scaling relation between X-ray count rate, redshift, and mass after calibration with weak-lensing shear measurements. Combined, these imply a total mass of $\sim$1.22 $\times 10^{15}\,M_\odot$, significantly higher than the SZ- and luminosity-based estimates. Our hydrostatic $\beta$-model measurements yield $M_{500} = (4.05^{+0.89}_{-0.79}) \times 10^{14}\,M_\odot$ for PSZ-E and $M_{500} = (2.44^{+0.57}_{-0.50}) \times 10^{14}\,M_\odot$ for PSZ-W. These values are lower than the weak-lensing-calibrated results of A. Liu et al. (2024), consistent with the expected hydrostatic mass bias in dynamically disturbed systems, but broadly compatible with the total system masses inferred from SZ and X-ray scaling relations.

### 4.2. X-Ray Morphology Indicators

As a first step, we calculated $w$ to measure the dynamic distortion level of each subcluster, which therefore quantifies any global asymmetry and gas disturbance. We obtain $w_{\rm PSZ\text{-}E} = 0.016 \pm 0.003$ and $w_{\rm PSZ\text{-}W} = 0.039 \pm 0.005$. $w_{\rm PSZ\text{-}E}$ is slightly above the empirical "relaxed" threshold of $w \lesssim 0.012$ (e.g., R. Cassano et al. 2010), while $w_{\rm PSZ\text{-}W}$ is well above it. PSZ-E is thus somewhat distorted, while the even larger value for PSZ-W points to a moderately disturbed dynamical core, consistent with ongoing merger activity and marked substructure.

We then computed three different power ratios for each subcluster to characterize their dynamical states. The first power ratio, $P_2/P_0$, is primarily sensitive to ellipticity in the ICM. We found $P_2/P_0 = (7.062 \pm 0.908) \times 10^{-6}$ for PSZ-E and $(2.330 \pm 0.199) \times 10^{-5}$ for PSZ-W, indicating that both subclusters exhibit significant ellipticity, with PSZ-W being more elongated. The presence of the X-ray bridge causes such significant ellipticities. We then measured the $P_3/P_0$ ratio, which traces asymmetries in the X-ray SB. We obtained $P_3/P_0 = (2.851 \pm 0.810) \times 10^{-6}$ for PSZ-E and $(1.204 \pm 0.576) \times 10^{-7}$ for PSZ-W. These values are typical of dynamically disturbed systems (R. Cassano et al. 2010), with PSZ-E appearing more asymmetric than PSZ-W. Finally, we computed the $P_4/P_0$ ratio, which is sensitive to small-scale structures and clumpiness within the ICM. For PSZ-E, we found $P_4/P_0 = (1.228 \pm 0.414) \times 10^{-7}$, while for PSZ-W we obtained a significantly lower value of $(6.231 \pm 8.333) \times 10^{-9}$. The relatively elevated $P_4/P_0$ in PSZ-E suggests the presence of small-scale substructures or clumpiness, possibly associated with recent merger activity. In contrast, the low value and large uncertainty in PSZ-W indicate the absence of clumpiness in the ICM distribution.

We measured two concentration indices, which target different spatial regimes ($c_{40}/c_{400}$ and $c_{100}/c_{500}$). On the inner 40–400 kpc scale, the J. S. Santos et al. (2008) criterion ($c_{40}/c_{400} > 0.075$) classifies PSZ-E ($c_{40}/c_{400} = 0.124 \pm 0.001$) as a moderate cool core, whereas PSZ-W ($c_{40}/c_{400} = 0.062 \pm 0.001$) falls below the cool-core limit. On the broader 100–500 kpc scale, however, both subclusters exceed the R. Cassano et al. (2010) nonrelaxed threshold ($c_{100}/c_{500} < 0.20$), with $c_{100}/c_{500} = 0.274 \pm 0.001$ for PSZ-E and $c_{100}/c_{500} = 0.213 \pm 0.002$ for PSZ-W, indicating a cool-core nature.

Here, we also compare our X-ray concentration parameter and centroid-shift estimates with the morphology thresholds defined by L. Lovisari et al. (2017), who calibrated these metrics using clusters in the Planck Early SZ sample. L. Lovisari et al. (2017) identify relaxed systems as those with a centroid shift $w < 0.021$ and a concentration $c > 0.15$. For PSZ2 G279.79+39.09, PSZ-E satisfies the concentration requirement ($c_{100/500} = 0.274 \pm 0.001$) but exhibits a borderline centroid shift ($w = 0.016 \pm 0.003$), placing it just above the relaxed threshold. PSZ-W, in contrast, shows both a large centroid shift ($w = 0.039 \pm 0.005$) and a lower concentration ($c_{100/500} = 0.213 \pm 0.002$), consistent with a disturbed system. These values place PSZ-E near the upper edge of the relaxed regime, while PSZ-W clearly occupies the disturbed, merging side of the L. Lovisari et al. (2017) morphological parameter space.

Table 1 summarizes three standard morphology metrics, the concentration index ($c$), centroid shift ($w$), and power-ratio moments ($P_m/P_0$). Taken together with the large centroid shifts and the elevated power ratios, we conclude that both halos are dynamically disturbed; PSZ-E still retains a residual cool-core signature on $\lesssim$40 kpc scales, while PSZ-W is borderline even by the broader $c_{100}/c_{500}$ metric. In other words, the merger has not yet erased the compact core of PSZ-E but has already disrupted its larger-scale symmetry, whereas PSZ-W appears non–cool core by either definition and exhibits the stronger global disturbance.

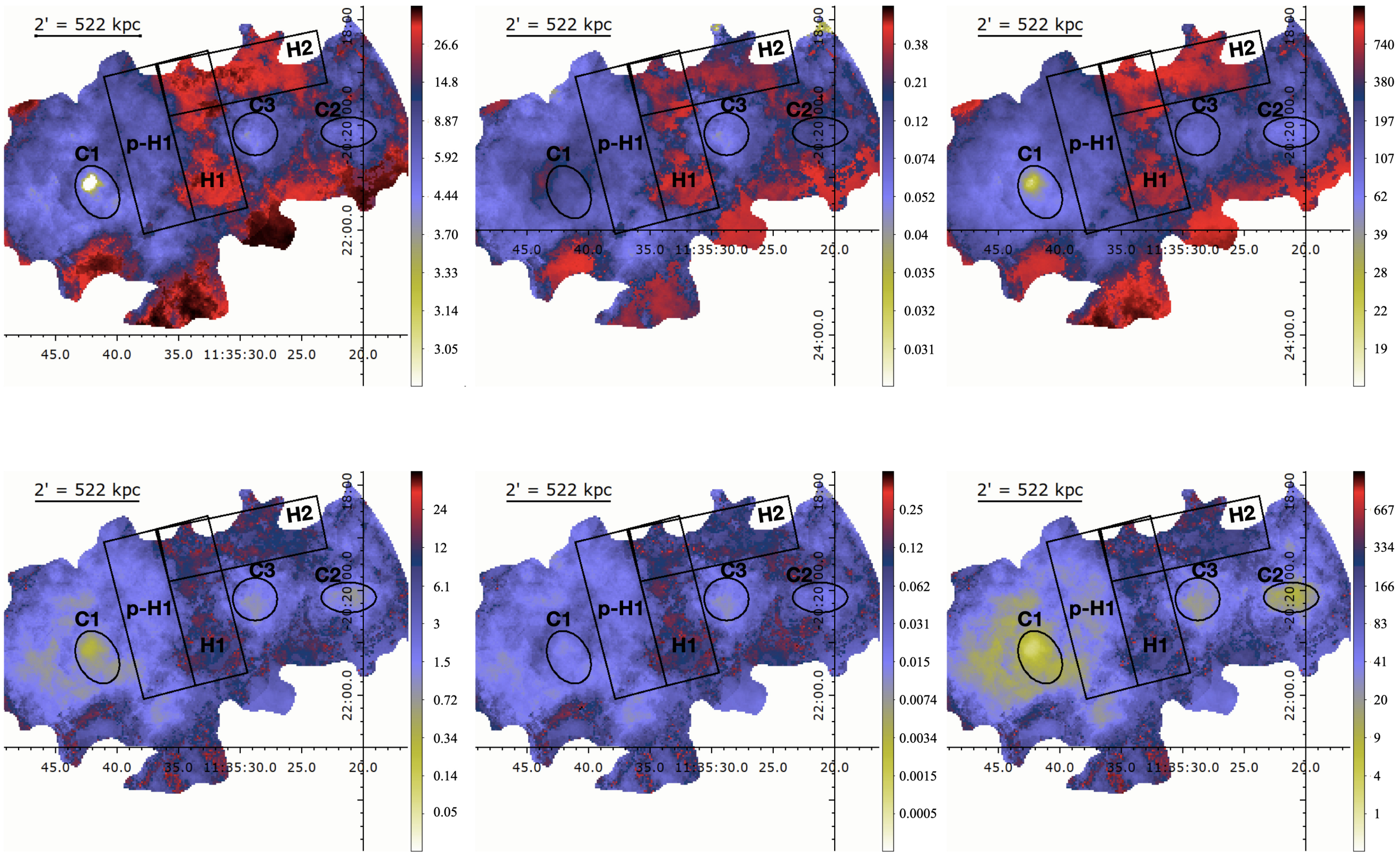

**Figure 3.** Top left: X-ray temperature map obtained using ACB and generating spectral templates via APEC model in XSPEC. Top middle: the pressure map. Top right: the entropy map. Bottom left: the temperature error map. Bottom middle: the pressure error map. Bottom right: the entropy error map. The selected cold and hot regions for spectral fitting are shown for visual aid. Due to low-count statistics in the bridge region, some spaxels yield significantly high values with substantial errors.

### 4.3. The Thermodynamic Maps

In Figure 3, we present projected temperature, pressure, and entropy maps of PSZ2 G279.79+39.09, together with their fractional-error maps. The thermodynamic structure shows that only the eastern subcluster (C1) exhibits a statistically significant cooler core relative to its own mean temperature. C1 has a temperature of $kT = 3.63^{+0.31}_{-0.27}$ keV, well below the cluster-wide median of $\sim$5.36 $\pm$ 0.14 keV, along with correspondingly low entropy and elevated central density, consistent with a surviving, pressure-confined core.

For the western subcluster, PSZ-W, however, the putative core region C2 has $kT = 5.10^{+0.76}_{-0.65}$ keV, which is statistically consistent with the global average of the system within the quoted uncertainties. Therefore, C2 cannot be classified as a cool core based on our data. Similarly, the western tail region C3 ($kT = 4.34^{+1.37}_{-1.09}$ keV) shows temperature, entropy, and pressure values that are consistent—within the large uncertainties—with the cluster-wide averages. Because of this, we do not claim C3 as a definitively cool, low-entropy, or low-pressure feature; we treat it instead as a tentative low-temperature substructure whose interpretation remains uncertain given the data quality.

We also identify two regions, H1 and H2, which appear hotter in the raw temperature map, but their temperatures are not significantly higher than neighboring values once statistical uncertainties are taken into account. H1 has $kT = 7.27^{+1.66}_{-1.26}$ keV compared to the control region P-H1 with $kT = 4.93^{+0.47}_{-0.40}$ keV, a difference that remains below the 1.5$\sigma$ level. A similar marginal enhancement is seen in H2. Although these temperature contrasts are not statistically significant, we may still formally evaluate the implied temperature jump to estimate the properties of a possible weak shock. Applying the Rankine–Hugoniot temperature jump condition to the nominal values yields a Mach number of $M = 1.48^{+0.30}_{-0.28}$, which corresponds to a sound speed of 11450 $\pm$ 50 km s$^{-1}$ and a shock velocity of $v_{\rm shock} = 1695^{+316}_{-300}$ km s$^{-1}$ (see T. Caglar 2018). As these features are detected at low significance, we describe H1 and H2 only as possible sites of shock-heated gas, and note that deeper X-ray observations will be required to confirm their nature.

### 4.4. The Merger Picture

In order to fully capture merger physics, we compare our results with numerical merger simulations (J. A. ZuHone 2011; J. A. ZuHone et al. 2018). In Table 2, we present our $kT$, $P$, and $K$ measurements for our control regions. The preshock

**Table 2**
Best-fit Spectral Parameters for Each Extraction Region

| Region | $kT$ (keV) | $Z$ ($Z_\odot$) | $n_e$ ($cm^{-3}$) | $P$ (keV $cm^{-3}$) | $K$ (keV $cm^2$) | $\chi^2$/DOF[1] |
|---|---|---|---|---|---|---|
| C1 | $3.63^{+0.31}_{-0.27}$ | $0.52^{+0.20}_{-0.18}$ | $(1.73 \pm 0.04) \times 10^{-2}$ | $(6.07 \pm 0.54) \times 10^{-2}$ | $52.6 \pm 4.5$ | 150.15/153 |
| C2 | $5.10^{+0.76}_{-0.65}$ | $0.42^{+0.32}_{-0.26}$ | $(2.85 \pm 0.05) \times 10^{-2}$ | $(1.45 \pm 0.20) \times 10^{-1}$ | $54.6 \pm 7.6$ | 135.72/118 |
| C3 | $4.34^{+1.37}_{-1.09}$ | 0.3(fixed) | $(9.05 \pm 0.49) \times 10^{-3}$ | $(3.91 \pm 1.10) \times 10^{-2}$ | $99.8 \pm 27.1$ | 124.41/139 |
| P-H1 | $4.93^{+0.47}_{-0.40}$ | $0.15^{+0.10}_{-0.09}$ | $(3.00 \pm 0.05) \times 10^{-3}$ | $(1.58 \pm 0.05) \times 10^{-2}$ | $254 \pm 7.3$ | 283.99/336 |
| H1 | $7.27^{+1.66}_{-1.26}$ | 0.3(fixed) | $(2.39 \pm 0.06) \times 10^{-3}$ | $(1.73 \pm 0.35) \times 10^{-2}$ | $406 \pm 79$ | 248.07/340 |
| H2 | $6.13^{+0.91}_{-0.70}$ | 0.3(fixed) | $(2.89 \pm 0.05) \times 10^{-3}$ | $(1.77 \pm 0.22) \times 10^{-2}$ | $303 \pm 39$ | 255.60/330 |

**Note.**
[1] Chi-squared divided by the degrees of freedom (DOF).

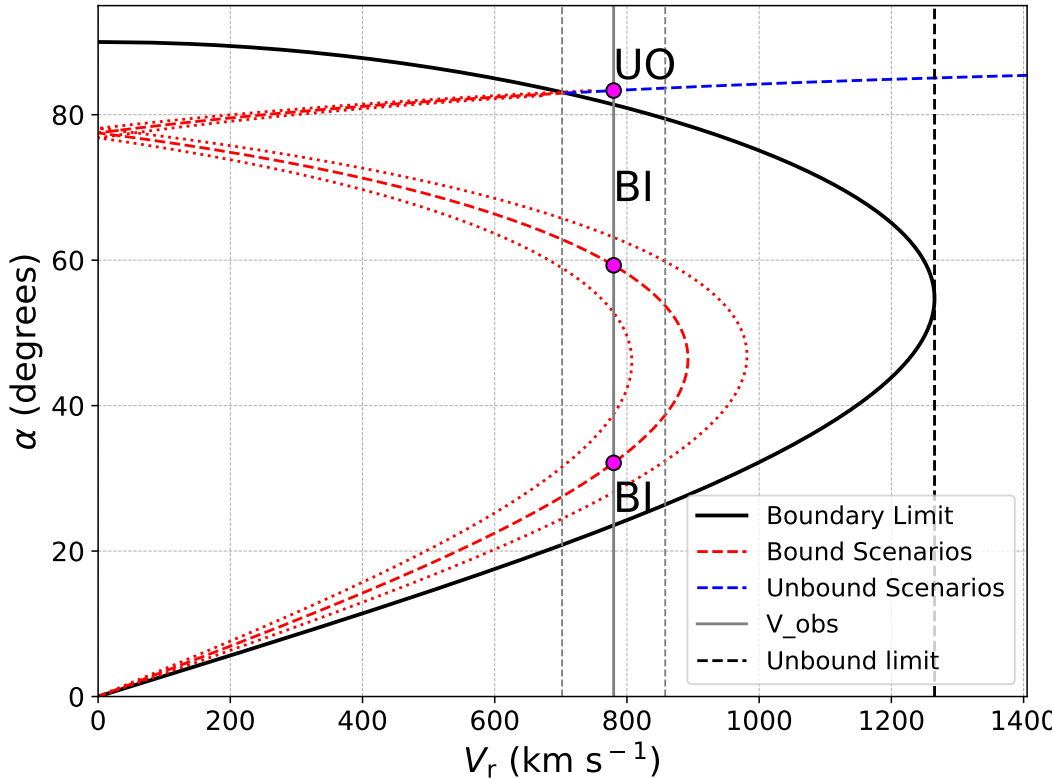

**Figure 4.** The projection angle ($\alpha$) as a function of the relative projected radial velocity difference ($V_r$) of our subclusters assuming the zero-angular-momentum head-on merger case. The black curve represents the limit of bound solutions due to the Newtonian criterion. The red and the blue dashed lines represent the bound and the unbound solutions, respectively. Three magenta points represent the possible two BI and one UO solutions. The red dotted line corresponds to the 68% confidence ranges. Using the photometric redshifts from M. Kluge et al. (2024), the radial velocity difference between the two BCGs is estimated to be $\sim$780 $\pm$ 2160 km $s^{-1}$, indicating that the redshift difference is not statistically significant. Nevertheless, to perform a simple Newtonian binding criterion test, we assume a nominal 10% uncertainty on the radial velocity difference (see the vertical gray solid and dashed lines) to explore the potential dynamical configuration of the system. A detailed spectroscopic follow-up campaign will be required to accurately determine the true velocity difference and dynamical state of the two subclusters.

bridge gas (P-H1) is still relatively cool and only moderately entropic, whereas the adjoining H1 region lies downstream of the weak $\mathcal{M}\sim1.5$ shock front, exhibiting the expected jump to higher temperature, pressure and entropy. The off-axis equal-mass merger simulations by J. A. ZuHone (2011) and J. A. ZuHone et al. (2018) predict that, before and shortly after the first core passage, the two cluster centers should still harbor low-entropy and low-temperature cores ($K \lesssim 100$ keV $cm^2$), while the gas stripped and shock heated in the intercluster region reaches a moderate-entropy bridge of $K \sim$ (250–500) keV $cm^2$.

Our X-ray measurements follow this pattern closely: each subcluster core exhibits $K_{\rm core} \sim 50$ keV $cm^2$, and the intercluster bridge reaches a much higher entropy of $K_{\rm bridge} \sim 410$ keV $cm^2$. This agreement supports an interpretation in which the subclusters are still approaching, observed after weak ($\mathcal{M} \sim 1.4$) shocks have heated the bridge gas but before the cool cores have been disrupted or fully mixed.

**Table 3**
Representative Bound-incoming and Unbound-outgoing Head-on Orbital Solutions Obtained from the Geometric Intersection Method

| Scenario | $\chi$ (rad) | $\alpha$ (deg) | $R_{\rm m}$ (Mpc) | $R_{\rm 3D}$ (Mpc) | $V_{\rm 3D}$ (km $s^{-1}$) | $P$ (%) |
|---|---|---|---|---|---|---|
| BI | 4.49 | 59.29 | 4.29 | 2.63 | 907.1 | 34.6 |
| BI | 4.94 | 32.13 | 4.07 | 1.59 | 1466.5 | 57.5 |
| UO | 1.00 | 83.32 | … | 11.55 | 785.2* | 7.9 |

**Note.** Probabilities $P$ are weighted by the isotropic orientation factor $\cos\alpha$. $\star$ is the asymptotic expansion velocity.

By examining the 1:1 and 1:3 mass merger simulations with 500 kpc impact parameters presented in J. A. ZuHone et al. (2018), together with the entropy evolution predicted by J. A. ZuHone et al. (2011), we find clear similarities with the observed properties derived from our X-ray analysis. In particular, the simulated systems reach projected separations comparable to PSZ2 G279.79+39.09 at times $\lesssim 0.5$ Gyr before or after pericenter passage. While the entropy structure and overall ICM morphology of PSZ-E and PSZ-W slightly favor a prepericenter (early stage) configuration, the available data do not exclude a postpericenter scenario, as both evolutionary phases remain consistent with idealized binary-merger simulations.

In the Appendix, we present our calculations for the Newtonian binding criterion for both the zero-angular-momentum (head-on) and nonzero-angular-momentum (off-axis) cases. Even though a fully off-axis geometry introduces deviations from the assumptions of the classical two-body model, the Newtonian approach still provides a useful first-order diagnostic for evaluating whether the observed configuration is dynamically consistent with bound or unbound orbital trajectories.

For the zero-angular-momentum (head-on) case, the allowed geometric intersections yield two physically meaningful solutions: a bound-incoming (BI) configuration and an unbound-outgoing (UO) hyperbolic configuration. The BI branch, which carries the largest probability weight (57.5%), corresponds to a first-infall trajectory with a three-dimensional velocity of $V_{\rm 3D} \approx 1466$ km $s^{-1}$ and a separation of $R_{\rm 3D} \approx 1.59$ Mpc. The second BI solution contributes an additional 34.6% probability at a lower infall velocity ($V_{\rm 3D} \approx 907$ km $s^{-1}$) and a larger separation ($R_{\rm 3D} \approx 2.63$ Mpc).

A third solution, accounting for only 7.9% of the posterior weight, corresponds to a UO hyperbolic encounter with $V_{\rm 3D} \approx 785$ km $s^{-1}$ and a large three-dimensional separation of

**Table 4**
Statistical Summary of Off-axis Orbital Solutions

| Statistic | $V_{3D,BI}$ (km s$^{-1}$) | $R_{3D,BI}$ (Mpc) | $f_{t,in}$ | $V_{3D,BO}$ (km s$^{-1}$) | $R_{3D,BO}$ (Mpc) | $f_{t,out}$ | $V_{3D,unb}$ (km s$^{-1}$) | $R_{3D,unb}$ (Mpc) | $f_{t,unbound}$ |
|---|---|---|---|---|---|---|---|---|---|
| Mode | 1147.9 | 1.722 | 0.950 | 1068.1 | 1.924 | 0.617 | 2253.2 | 2.044 | 0.503 |
| Mean | 1320.4 | 1.748 | 0.552 | 997.9 | 2.568 | 0.611 | 2058.7 | 2.099 | 0.435 |
| Median | 1277.1 | 1.733 | 0.584 | 1008.3 | 2.261 | 0.623 | 2103.7 | 1.703 | 0.438 |
| Std. dev. | 172.8 | 0.068 | 0.292 | 75.2 | 0.860 | 0.233 | 265.1 | 2.295 | 0.180 |

**Note.** The orbital phase variable $f_t$ denotes the dimensionless time along the orbit; $f_{t,\rm in}$ for BI; $f_{t,\rm out}$ for BO; and $f_{t,\rm unbound}$ is represented by the absolute hyperbolic anomaly $|F_{\rm hyp}|$ for unbound solutions.

$R_{3D} \approx 11.55$ Mpc. In the left panel of Figure 4 and in Table 3, we summarize these three geometrically allowed solutions obtained from the projection-angle and line-of-sight velocity constraints.

In a real off-axis, postpericenter scenario, the cluster may never return exactly along the same direction—its separation and line-of-sight velocity evolve differently. In practice, bound-outgoing (BO) scenarios for head-on mergers end up with large projection angles ($\alpha$) and low probability values that probably do not match the actual geometry for off-axis mergers. This behavior arises because the head-on model enforces zero angular momentum ($L = 0$), forcing all motion to lie along the collision axis. Once the subclusters pass pericenter, any real transverse velocity component cannot be represented, so the model compensates by inflating the inferred 3D separation and reducing the relative velocity to reproduce the same projected observables. As a result, the BO branch in head-on configurations systematically yields unrealistically large $R_{3D}$ values.

To explore all dynamically allowed off-axis configurations, we sampled the full space of bound and unbound orbital solutions consistent with the observed projected separation and line-of-sight velocity. The resulting posterior probabilities assign 49% to BI (first-infall), 35% to BO (postpericenter), and 16% to unbound hyperbolic encounters. The first-infall (BI) solution therefore represents the most likely scenario. These solutions yield characteristic three-dimensional parameters of $V_{3D} \approx 1277$ km s$^{-1}$ and $R_{3D} \approx 1.73$ Mpc (median values), consistent with a gravitationally bound system approaching its first core passage (see Table 4).

For bound orbits, we also compute the dimensionless orbital phase parameter $f_t$, which expresses where the system lies along its bound trajectory: $f_t = 0$ corresponds to pericenter and $f_t = 1$ corresponds to apocenter. In this framework, first-infall solutions have $0 < f_t < 1$ as the system moves from apocenter toward pericenter, while postpericenter solutions have $0 < f_t < 1$ as the system moves away from pericenter toward apocenter. The median values—$f_{t,\rm in} \simeq 0.58$ for the first-infall branch and $f_{t,\rm out} \simeq 0.62$ for the postpericenter branch—place both families comfortably within their expected orbital domains, with the first-infall values indicating a system more than halfway from apocenter toward its first core encounter (see Table 4).

The BO branch, interpreted as postpericenter trajectories, occupies a smaller fraction of the posterior and favors larger three-dimensional separations, yielding median values of $V_{3D} \approx 1008$ km s$^{-1}$ and $R_{3D} \approx 2.26$ Mpc. The unbound solutions form only a minor component of the distribution and are characterized by substantially higher velocities ($V_{3D} \approx 2104$ km s$^{-1}$, median) and weaker geometric compatibility.

Taken together, the posterior distributions indicate that the system is most likely on its first infall, prior to any core passage, although a postpericenter state remains dynamically possible. We do not attempt to extend the interpretation to subsequent orbital cycles: after the first apocenter, the ICM becomes highly disrupted and the projected geometry no longer predictable, making later infall phases intrinsically unconstrained by this approach. The present analysis therefore distinguishes only between first-infall and first-outgoing phases, with the former moderately favored by the orbital solutions.

## 5. Conclusion

In this work, we present the detailed X-ray analysis of a binary cluster: PSZ2 G279.79+39.09. We present evidence that the system is experiencing an off-axis merger, and is observed shortly before or after pericenter passage. Here, we list our main findings as follows.

1. The X-ray SB map of PSZ2 G279.79+39.09 displays a bimodal morphology, with the emission elongated along the east–west axis extending to a projected radius of $\sim$1.3 Mpc.
2. We find a mean $kT$ of $5.36 \pm 0.15$ keV and a $Z$ of $0.34 \pm 0.04\, Z_\odot$ for PSZ-E and $5.44 \pm 0.65$ keV and $0.29 \pm 0.07\, Z_\odot$, respectively, for PSZ-W.
3. Based on X-ray morphology parameters, we compute large centroid shifts and elevated power ratios. We conclude that both halos are dynamically disturbed; PSZ-E still retains a residual moderate cool core, while PSZ-W is borderline by the broader $c_{100}/c_{500}$ metric.
4. We detect a high-temperature, pressure, and entropy region in between the subclusters, indicating shock heating in the ICM. Applying the Rankine–Hugoniot temperature jump condition, we calculate a Mach number ($M$) of $1.48^{+0.30}_{-0.28}$ indicating a weak merger shock propagating through the ICM with a shock velocity of $1695^{+316}_{-300}$[ km s$^{-1}$.
5. We measure $\log M_{500} = (4.05^{+0.89}_{-0.79}) \times 10^{14}\, M_\odot$ for PSZ-E and $(2.44^{+0.57}_{-0.50}) \times 10^{14}\, M_\odot$ for PSZ-W. Their combined mass is therefore $\sim 10^{15}\, M_\odot$, with a mass ratio of 1:1.7, placing this system among the most massive cluster mergers known.
6. Taken together with the off-axis Newtonian orbital analysis, our results indicate that a first-infall, prepericenter stage of an off-axis binary merger is the most likely scenario.

Our X-ray analysis favors a coherent picture in which the two subclusters are close to pericenter, colliding with a nonzero impact parameter. While we cannot firmly determine the exact merger phase (pre- or postcore passage), the data slightly favor the scenario in which the two systems have not yet experienced core passage. Nevertheless, several key steps are still required to confirm the true dynamical stage of the merger. First, optical spectroscopy of cluster members is essential to map the line-of-sight velocity field and confirm the dynamical state of the two subclusters. Second, deeper X-ray exposures will allow fully deprojected temperature, density, pressure, and entropy profiles, reducing the current uncertainties in the bridge region. Finally, dedicated hydrodynamic simulations matched to the observed mass ratio, impact parameter, and viewing angle will provide the quantitative framework needed to translate these thermodynamic maps into a detailed merger timeline.

## Acknowledgments

We thank the anonymous referee for a careful and constructive review, which helped us to improve the clarity, robustness, and presentation of this work. Support for J.A.Z. was provided by the Chandra X-ray Observatory Center, which is operated by the Smithsonian Astrophysical Observatory for and on behalf of NASA under contract NAS8-03060.

The XMM-Newton data utilized in this paper are publicly available in the XMM-Newton data archive at https://nxsa.esac.esa.int/nxsa-web/#search. This paper employs a single Chandra dataset, obtained by the Chandra X-ray Observatory, contained in the Chandra Data Collection (CDC): DOI: 10.25574/cdc.533.

*Facilities:* CXO, XMM.

*Software:* CIAO (A. Fruscione et al. 2006), SAS (C. Gabriel et al. 2004), ESAS (S. L. Snowden et al. 2008), XSPEC (K. A. Arnaud 1996), Astropy (Astropy Collaboration et al. 2013, 2018), NumPy (C. R. Harris et al. 2020), SciPy (P. Virtanen et al. 2020), Matplotlib (J. D. Hunter 2007), SAOImage DS9 (W. A. Joye & E. Mandel 2003).

## Appendix
## The Newtonian Binding Criterion

### A.1. Head-on (Zero-angular-momentum) Case

As we assume that the two systems are bound and falling toward each other with a certain infalling angle, the Newtonian criterion for gravitational binding of two-body systems can be solved from the following equation (T. C. Beers et al. 1982, 1991):

$$V_r^2 R_p \leqslant 2GM \sin^2\alpha \cos\alpha, \tag{A1}$$

where $V_r$ is the radial velocity difference, $G$ is the gravitational constant, $R_p$ is the projected separation, and $\alpha$ is the projection angle. This parametric equation solves the projection angle $\alpha$ for each radial velocity difference $V_r$. The radial velocity and the projected separation are related to the projection angle of the system:

$$V_r = V \sin\alpha,\ R_p = R\cos\alpha, \tag{A2}$$

where $V$ and $R$ are the three-dimensional velocity difference and separation of the two-body system in the field of sky, respectively. The parametric motion equation of the bound system can be computed from the following equations:

$$t = \left(\frac{R_m^3}{8GM}\right)^{1/2}(\chi - \sin\chi), \tag{A3}$$

$$R = \frac{R_m}{2}(1 - \cos\chi), \tag{A4}$$

$$V = \left(\frac{2GM}{R_m}\right)^{1/2}\frac{\sin\chi}{(1 - \cos\chi)}, \tag{A5}$$

where $R$ is the separation at time $t$, $R_m$ is the separation at the maximum expansion, $\chi$ is the development angle, $M$ is the total mass of the system.

For gravitationally unbound systems:

$$V = V_\infty \frac{\sinh\chi}{(\cosh\chi - 1)}, \tag{A6}$$

where $V_\infty$ is the expansion velocity at the asymptotic limit.

The relative probabilities of the solutions can be estimated using the formula:

$$p_j = \int_{\alpha_{\mathrm{inf},j}}^{\alpha_{\mathrm{sup},j}} \cos\alpha\, d\alpha, \tag{A7}$$

where $j$ represents the $j$th solution of possible binding scenarios. The probabilities were then normalized to $P_j = p_j/(\sum_j p_k)$ in our calculations. However, this method is only applicable to head-on mergers, for which we assume angular momentum ($L$) to be zero (T. C. Beers et al. 1982, 1991). In contrast, off-axis mergers automatically require a nonzero angular momentum. In our updated method, we also allow for off-axis (nonzero-angular-momentum) orbits by introducing directly the triple ($e$, $\alpha$, and $\chi$) in a Monte Carlo scan.

### A.2. Off-axis (Nonzero-angular-momentum) Case

To model the full range of dynamically allowed off-axis orbits, we sample the Keplerian parameter space in eccentricity, orientation, and orbital phase. We draw the eccentricity $e$ uniformly in the interval (0.01, 2.0), so that values with $e < 1$ correspond to bound (elliptic) solutions, while $e > 1$ correspond to unbound (hyperbolic) trajectories. The orientation is described by the projection angle $\alpha \in [0, \pi/2]$, measured from the line of sight, drawn with the isotropic prior $P(\alpha) \propto \sin\alpha$ (equivalently, $\cos\alpha$ is uniform in [0, 1]). For bound orbits we draw the eccentric anomaly $E$ uniformly in $[-\pi, \pi]$, which spans the first inbound branch ($-\pi < E < 0$) and the first outbound branch ($0 < E < \pi$). For unbound solutions, we draw the hyperbolic anomaly $F$ uniformly from $[-F_{\max}, F_{\max}]$ with $F_{\max} = 2$. A total of $5 \times 10^8$ random triples ($e$, $\alpha$, and $E/F$) are generated.

For bound orbits ($e < 1$), the pericenter distance is linked to the observed projected separation via $q = a(1 - e) \simeq R_p$, giving

$$a_\mathrm{b} = \frac{R_p}{1 - e}. \tag{A8}$$

The three-dimensional separation and velocities are

$$r_\mathrm{b}(E) = a_\mathrm{b}(1 - e\cos E), \tag{A9}$$

$$\dot{r}_\mathrm{b}(E) = \sqrt{\frac{GM}{a_\mathrm{b}}}\,\frac{e\sin E}{1 - e\cos E}, \tag{A10}$$

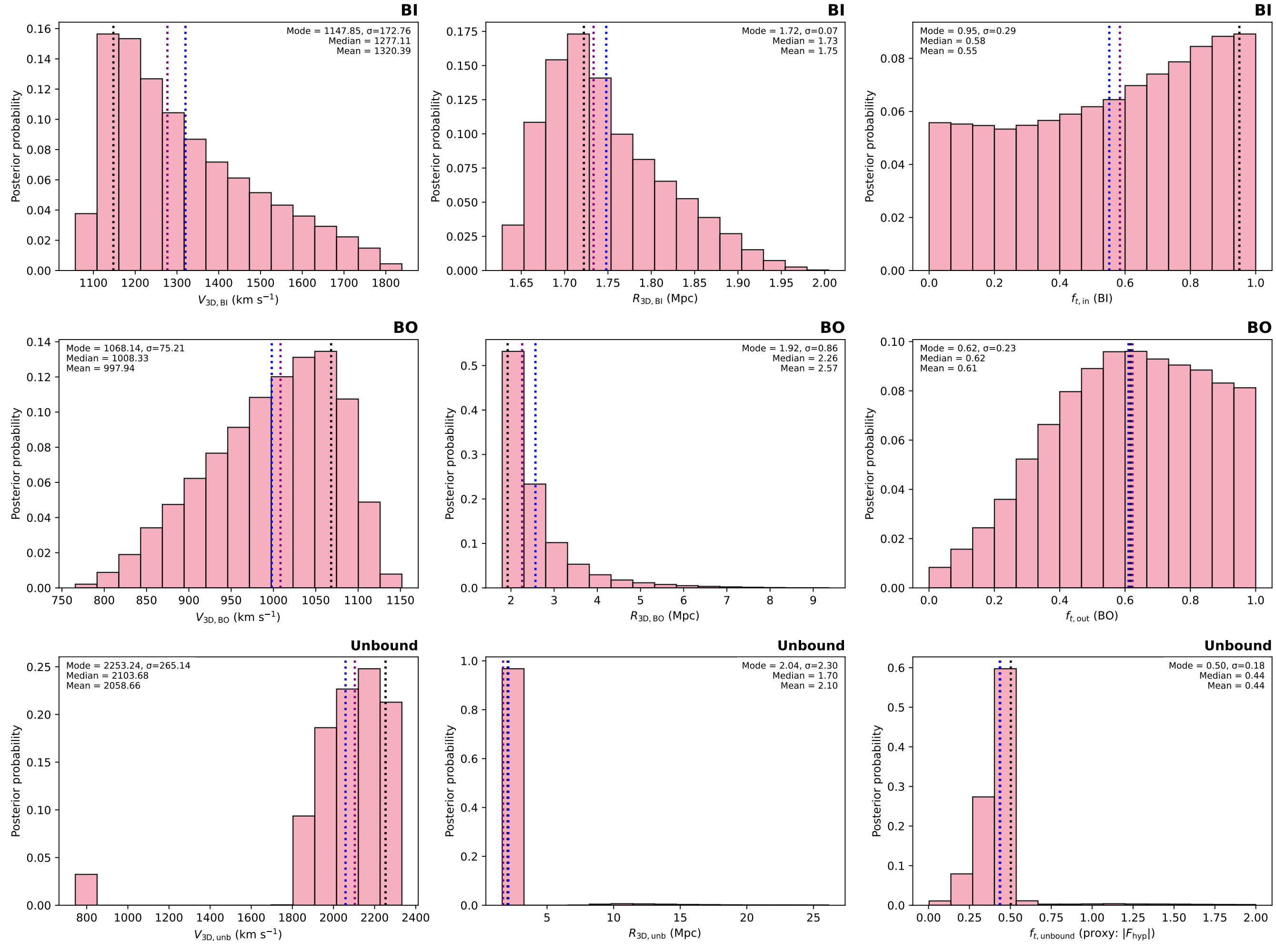

**Figure 5.** Posterior probability distributions of the three-dimensional infall velocities (left), three-dimensional separations (middle), and orbital phase parameters (right) for all dynamically allowed off-axis orbital families. The orbital phase variable $f_t$ represents the dimensionless time along the orbit; $f_{t,\mathrm{in}}$ measures the fractional time from apocenter to pericenter for the BI (first-infall) solutions; $f_{t,\mathrm{out}}$ measures the fractional time from pericenter to apocenter for the BO (postpericenter) solutions; and $f_{t,\mathrm{unbound}}$ is shown via the absolute hyperbolic anomaly $|F_{\mathrm{hyp}}|$ for unbound trajectories. Top row: BI solutions. Middle row: BO solutions. Bottom row: unbound hyperbolic solutions. Vertical black, purple, and blue lines mark the weighted mode, median, and mean of each distribution, respectively.

$$v_{\perp,\mathrm{b}}(E) = \sqrt{\frac{GM}{a_{\mathrm{b}}}}\,\frac{\sqrt{1-e^2}}{1-e\cos E}. \tag{A11}$$

For unbound orbits ($e > 1$), the pericenter condition $q = a(e-1) \simeq R_p$ gives

$$a_{\mathrm{u}} = \frac{R_p}{e-1}, \tag{A12}$$

with

$$r_{\mathrm{u}}(F) = a_{\mathrm{u}}(e\cosh F - 1), \tag{A13}$$

$$\dot{r}_{\mathrm{u}}(F) = \sqrt{\frac{GM}{a_{\mathrm{u}}}}\,\frac{e\sinh F}{e\cosh F - 1}, \tag{A14}$$

$$v_{\perp,\mathrm{u}}(F) = \sqrt{\frac{GM}{a_{\mathrm{u}}}}\,\frac{\sqrt{e^2-1}}{e\cosh F - 1}. \tag{A15}$$

Each draw is projected onto the sky using

$$R' = r\sin\alpha,\;\; V' = \dot{r}\cos\alpha + v_{\perp}\sin\alpha. \tag{A16}$$

The pair ($R'$, $V'$) is accepted as a valid solution if both match the observed projected separation $R_p$ and line-of-sight velocity $V_r$ within a common 5% tolerance:

$$\left|\frac{R' - R_p}{R_p}\right| < 0.05, \quad \left|\frac{V' - V_r}{V_r}\right| < 0.05.$$

Each accepted solution is assigned a geometric probability weight proportional to $\cos\alpha$, consistent with an isotropic orientation distribution. This yields 509,951 valid off-axis solutions spanning 52% BI (first-infall), 30% BO (postpericenter), and 18% unbound hyperbolic branches.

To characterize the orbital phase, we use the Keplerian mean anomaly $M$, which increases linearly with time. For BI solutions, corresponding to $-\pi < E < 0$, the mean anomaly runs from $M = -\pi$ at apocenter to $M = 0$ at pericenter, and we define

$$f_{t,\mathrm{in}} = \frac{M+\pi}{\pi}, \tag{A17}$$

which increases from zero (apocenter) to one (pericenter). For BO solutions with $0 < E < \pi$, the motion proceeds from pericenter ($M = 0$) to apocenter ($M = \pi$), and we define

$$f_{t,\text{out}} = \frac{M}{\pi}. \quad \text{(A18)}$$

Unbound trajectories do not complete an orbital cycle, so we use the absolute value of the hyperbolic anomaly as a phase-like coordinate:

$$f_{t,\text{unbound}} \equiv |F|, \quad \text{(A19)}$$

which measures the temporal distance from pericenter ($F = 0$) along the hyperbolic branch, with larger values corresponding to configurations farther from the closest approach. The posterior distributions of $V_{3\text{D}}$, $R_{3\text{D}}$, and the three phase indicators ($f_{t,\text{in}}$, $f_{t,\text{out}}$, and $f_{t,\text{unbound}}$) are shown in Figure 5.

## ORCID iDs

Sibel Döner https://orcid.org/0000-0001-7922-8961
Turgay Caglar https://orcid.org/0000-0002-9144-2255
Krista L. Smith https://orcid.org/0000-0001-5785-7038
Serap Ak https://orcid.org/0000-0002-0912-6019
Andrea Botteon https://orcid.org/0000-0002-9325-1567
M. Kiyami Erdim https://orcid.org/0000-0002-6770-5043
John A. ZuHone https://orcid.org/0000-0003-3175-2347

## References

Anders, E., & Grevesse, N. 1989, GeCoA, 53, 197
Arnaud, K. A. 1996, ASPC, 101, 17
Astropy Collaboration, Robitaille, T. P., Tollerud, E. J., et al. 2013, A&A, 558, A33
Astropy Collaboration, Price-Whelan, A. M., Sipőcz, B. M., et al. 2018, AJ, 156, 123
Beers, T. C., Geller, M. J., & Huchra, J. P. 1982, ApJ, 257, 23
Beers, T. C., Gebhardt, K., Forman, W., Huchra, J. P., & Jones, C. 1991, AJ, 102, 1581
Böhringer, H., & Werner, N. 2010, A&ARv, 18, 127
Botteon, A., Gastaldello, F., & Brunetti, G. 2018, MNRAS, 476, 5591
Botteon, A., Cassano, R., Eckert, D., et al. 2019, A&A, 630, A77
Botteon, A., van Weeren, R. J., Brunetti, G., et al. 2020, MNRAS, 499, L11
Botteon, A., van Weeren, R. J., Eckert, D., et al. 2024, A&A, 690, A222
Botteon, A., Caglar, T., Döner, S., van Weeren, R. J., & Smith, K. L. 2026, A&A, 705, A230
Bourdin, H., Mazzotta, P., Markevitch, M., Giacintucci, S., & Brunetti, G. 2013, ApJ, 764, 82
Buote, D. A., & Tsai, J. C. 1995, ApJ, 452, 522
Caglar, T. 2018, MNRAS, 475, 2870
Caglar, T., & Hudaverdi, M. 2017, MNRAS, 472, 2633
Cash, W. 1979, ApJ, 228, 939
Cassano, R., Ettori, S., Giacintucci, S., et al. 2010, ApJL, 721, L82
Cavaliere, A., & Fusco-Femiano, R. 1976, A&A, 49, 137
Ebeling, H., Qi, J., & Richard, J. 2017, MNRAS, 471, 3305
Eckert, D., Roncarelli, M., Ettori, S., et al. 2015, MNRAS, 447, 2198
Ettori, S., Morandi, A., Tozzi, P., et al. 2009, A&A, 501, 61
Feretti, L., Giovannini, G., Govoni, F., & Murgia, M. 2012, A&ARv, 20, 54
Forman, W. R. 2017, AAS/HEAD Meeting, 16, 105.17
Fruscione, A., McDowell, J. C., Allen, G. E., et al. 2006, SPIE, 6270, 62701V
Gabriel, C., Denby, M., Fyfe, D. J., et al. 2004, ASPC, 314, 759
Govoni, F., Orrù, E., Bonafede, A., et al. 2019, Sci, 364, 981
Gu, L., Akamatsu, H., Shimwell, T. W., et al. 2019, NatAs, 3, 838
Ha, J.-H., Ryu, D., & Kang, H. 2018, ApJ, 857, 26
Hallman, E. J., Alden, B., Rapetti, D., Datta, A., & Burns, J. O. 2018, ApJ, 859, 44
Harris, C. R., Millman, K. J., van der Walt, S. J., et al. 2020, Natur, 585, 357
Hunter, J. D. 2007, CSE, 9, 90
Joye, W. A., & Mandel, E. 2003, ASPC, 295, 489
Kalberla, P. M. W., Burton, W. B., Hartmann, D., et al. 2005, A&A, 440, 775
Kaya, H. I., Caglar, T., & Sert, H. 2019, MNRAS, 485, 4550
Klein, M., Hernández-Lang, D., Mohr, J. J., Bocquet, S., & Singh, A. 2023, MNRAS, 526, 3757
Kluge, M., Comparat, J., Liu, A., et al. 2024, A&A, 688, A210
Laganá, T. F., Martinet, N., Durret, F., et al. 2013, A&A, 555, A66
Lima Neto, G. B., Capelato, H. V., Sodré, L., Jr., & Proust, D. 2003, A&A, 398, 31
Liu, A., Bulbul, E., Kluge, M., et al. 2024, A&A, 683, A130
Lovisari, L., Forman, W. R., Jones, C., et al. 2017, ApJ, 846, 51
Machado, R. E. G., Laganá, T. F., Souza, G. S., et al. 2022, MNRAS, 515, 581
Mann, A. W., & Ebeling, H. 2012, MNRAS, 420, 2120
Markevitch, M., & Vikhlinin, A. 2007, PhR, 443, 1
McCarthy, I. G., Bower, R. G., Balogh, M. L., et al. 2007, MNRAS, 376, 497
Migkas, K., Sommer, M. W., Schrabback, T., et al. 2025, A&A, 694, A45
Mirakhor, M. S., Walker, S. A., & Runge, J. 2022, MNRAS, 509, 1109
Mohr, J. J., Evrard, A. E., Fabricant, D. G., & Geller, M. J. 1995, ApJ, 447, 8
Nevalainen, J., Markevitch, M., & Lumb, D. 2005, ApJ, 629, 172
Niemiec, A., Jauzac, M., Eckert, D., et al. 2023, MNRAS, 524, 2883
Ota, N., Onzuka, K., & Masai, K. 2013, PASJ, 65, 47
Pineau, F. X., Derriere, S., Motch, C., et al. 2017, A&A, 597, A89
Planck Collaboration, Ade, P. A. R., Aghanim, N., et al. 2014, A&A, 571, A29
Planck Collaboration, Ade, P. A. R., Aghanim, N., et al. 2016, A&A, 594, A27
Roediger, E., Brüggen, M., Simionescu, A., et al. 2011, MNRAS, 413, 2057
Santos, J. S., Rosati, P., Tozzi, P., et al. 2008, A&A, 483, 35
Sarazin, C. L. 2002, ASSL, 272, 1
Seppi, R., Comparat, J., Nandra, K., et al. 2023, A&A, 671, A57
Shi, X., Nagai, D., Aung, H., & Wetzel, A. 2020, MNRAS, 495, 784
Smith, R. K., Brickhouse, N. S., Liedahl, D. A., & Raymond, J. C. 2001, ApJL, 556, L91
Snowden, S. L., Mushotzky, R. F., Kuntz, K. D., & Davis, D. S. 2008, A&A, 478, 615
Stroe, A., Rajpurohit, K., Zhu, Z., et al. 2025, ApJ, 984, 24
Tarrío, P., Melin, J. B., & Arnaud, M. 2019, A&A, 626, A7
Toptun, V., Popesso, P., Marini, I., et al. 2025, A&A, 700, A167
Vazza, F., Eckert, D., Simionescu, A., Brüggen, M., & Ettori, S. 2013, MNRAS, 429, 799
Virtanen, P., Gommers, R., Oliphant, T. E., et al. 2020, NatMe, 17, 261
Xu, H., Makishima, K., Fukazawa, Y., et al. 1998, ApJ, 500, 738
ZuHone, J. A. 2011, ApJ, 728, 54
ZuHone, J. A., Markevitch, M., & Johnson, R. E. 2010, ApJ, 717, 908
ZuHone, J. A., Markevitch, M., & Lee, D. 2011, ApJ, 743, 16
ZuHone, J. A., Kowalik, K., Öhman, E., Lau, E., & Nagai, D. 2018, ApJS, 234, 4